# New grating compressor designs for XCELS and SEL-100 PW projects

**EFIM KHAZANOV**

*Gaponov-Grekhov Institute of Applied Physics of the Russian Academy of Sciences, Russia*
*efimkhazanov@gmail.com*

**Abstract:** The problem of optimizing the parameters of a laser pulse compressor consisting of four identical diffraction gratings is solved analytically. The goal of optimization is to obtain maximum pulse power, completely excluding both beam clipping on gratings and the appearance of spurious diffraction orders. The analysis is carried out in a general form for an out-of-plane compressor. Two particular "plane" cases attractive from a practical point of view are analyzed in more detail: a standard Treacy compressor (TC) and a compressor with an angle of incidence equal to the Littrow angle (LC). It is shown that in both cases the LC is superior to the TC. Specifically, for 160-cm diffraction gratings, optimal LC design enables 109 PW for XCELS and 115 PW for SEL-100 PW, while optimal TC design enables 86 PW for both projects.

## 1. Introduction

In high-power femtosecond lasers, the output pulse energy is limited by the pump pulse energy and the laser induced damage threshold of compressor diffraction gratings. The latter limitation is stronger in 100-PW laser projects [1-9], where Nd:glass laser pulses with an energy of about 10 kJ are used for pumping. The damage threshold of gratings by nanosecond pulses is much higher than by femtosecond ones [10]. Therefore, despite a less energy incident on the last grating than on the first one, the laser damage threshold of the last grating is of major importance. Thus, the maximum output energy $W$ is proportional to squared beam size $d$, threshold value of fluence $w_{th}$ (in the plane normal to the beam wave vector), reflection coefficient $R$ of the grating, and fill-factor $\eta$ taking into account fluence inhomogeneity in the beam:

$$W = R\eta w_{th} d^2. \qquad (1)$$

Here, we assume that the beam has a square cross section. Increasing $w_{th}$ and $R$ is a technological task, that is beyond the scope of this paper. The fill-factor $\eta$ depends on the energy and spectral properties of the spatial noise of the beam, in particular, on rms and effective spatial frequency [11]. Both these parameters can be significantly reduced by using an asymmetric compressor [5, 12-14] or a compressor with an out-of-plane geometry [15]. The purpose of this work is to search for the following compressor parameters: angle of incidence on the first grating $\alpha$, distance between the gratings along the normal $L$, and groove density $N$ that allow obtaining the maximum value of $W$. Bearing in mind that $w_{th}$ does not depend on the angle of incidence $\alpha$ on the grating [16, 17], we will assume that $w_{th}$, $R$, and $\eta$ are constants which do not depend on the compressor parameters. Thus, an optimal compressor design ($\alpha$, $L$, $N$) is a design that ensures a maximum value of $d^2$. Note that in the expression (1) $R$ is to the power of one rather than four, as laser induced damage restrictions are important only for the last grating.

The main restriction on increasing $d$ is the fact that on the second grating the beam size should not be larger than the grating length $L_g$. A standard compressor [18] consists of identical gratings, with the gratings of the first and second pair being antiparallel to each other, see Fig.

1a. We will further call such a compressor a Treacy compressor (TC). TC is used in the vast majority of high-power lasers [19]. The maximization of $d$ was considered in [20] in the $\omega_0 \gg \Omega$ approximation ($\omega_0$ is center frequency and $\Omega$ is bandwidth). For pulses with a duration less than 50 fs, this approximation is not accurate. However, in this case it can be readily shown that, for a given dispersion of a chirped pulse and given $N$, $d$ is proportional to $\cos\alpha$. Then, from (1) it can be found that for increasing $W$ it is necessary to decrease $\alpha$. However, the decrease in $\alpha$ makes decoupling impossible, i.e. the condition that the second grating must not overlap with the input beam cannot be fulfilled. It is obvious that decoupling is impossible if $\alpha \approx \alpha_L$, where $\alpha_L$ is the Littrow angle. As will be shown below for a general case, i.e. outside the $\omega_0 \gg \Omega$ approximation, at certain parameters an optimal compressor is TC with $\alpha < \alpha_L$.

For $\alpha \approx \alpha_L$, decoupling may be provided employing an out-of-plane compressor [21] that is used, for example, for spectral beam combining [22] and for compressing narrow-beam pulses [23]. In this work we propose to use an out-of-plane compressor for increasing output power by decreasing $\alpha$ down to $\alpha = \alpha_L$ and $\alpha < \alpha_L$ inclusive. Both multilayer dielectric [24] and gold gratings [16, 24] in the out-of-plane geometry may have a reflection coefficient $R$ almost the same as in the out-of-plane geometry. It is important to note that for $\alpha = \alpha_L$ the out-of-plane compressor "turns out" to be plane again (Fig. 1 b), which greatly simplifies its experimental implementation. Such a compressor will be referred to as LC. LC has a number of additional advantages [24], one of which is the use of multilayer dielectric gratings the reflection band of which rapidly narrows with increasing $(\alpha - \alpha_L)$, which makes them unfit for TC in wideband lasers [25]. An important issue of radiation polarization in the out-of-plane compressor was discussed in detail in Ref. [24].

Analytical expressions that allow finding the compressor parameters which provide maximum values of $d$ for both TC and LC will be obtained in Section 2. Optimal designs of both compressors for the XCELS project [4] will be discussed in detail in Section 3. An analogous optimization for the pulse parameters of the SEL-100 PW project [1, 3, 10] will be made in Section 4.

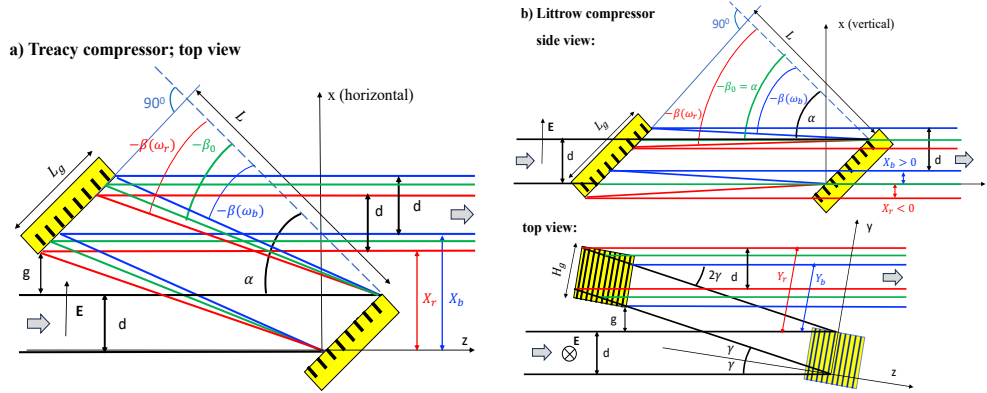

Fig. 1. TC (a) and LC (b). The second half of the compressor (third and fourth gratings) is absolutely symmetric to the first one, so it is not shown in the figure. The angle of reflection in the diffraction plane is $\beta < 0$, which explains the minus sign in the figure. The angle of reflection in the plane orthogonal to the diffraction plane is always equal to the angle of incidence $\gamma$.

## 2. Maximum beam size for TC and LC

We will first consider a general case of an out-of-plane compressor when the angles of incidence on the first grating in two planes $\gamma$ and $\alpha$ are arbitrary. TC (Fig. 1a) and LC (Fig. 1b) are its particular cases at $\gamma = 0$ and $\alpha = \alpha_L$, respectively. Note that, both in TC and LC, the gratings of the first and second pairs are antiparallel (mirror) to each other in the planes orthogonal to incident beam. The case of non-parallel gratings is considered, for example, in [26]. Maximum beam size will be determined using the following procedure. We choose the coordinate system $(x, y, z)$ as shown in Fig. 1: the $y$-axis is parallel to the direction of the grooves, and the $x$-axis in the $(x, z)$ diffraction plane is directed at an angle $\alpha$ to the surface of the grating. The coordinate origin coincides with the point of incidence of the beam on the first grating. Let us find the spectral phase $\Psi(\omega, k_x, k_y)$ accumulated in the beam on reflection from the first grating, propagation to the second grating, reflection from the second grating, and propagation to the $z = 0$ plane. The first derivatives of $\Psi$ with respect to $k_x, k_y$ up to the sign are equal to the beam coordinates $X(\omega)$ and $Y(\omega)$ in the $z = 0$ plane. These coordinates will allow, for geometric reasons, to determine maximum beam size $d$ depending on the parameters of the compressor and the input pulse. The expression for $\Psi$ is available in [18, 27]; in the chosen coordinate system it has the form:

$$\Psi(\omega, k_x, k_y) = Lk_{zx}\left(cos\theta + cos\left\{\alpha + atan\frac{k_x}{k_z}\right\}\right), \qquad (2)$$

where $k_{zx}^2 = \frac{\omega^2}{c^2} - k_y^2$, $k_z^2 = \frac{\omega^2}{c^2} - k_x^2 - k_y^2$, and $\theta$ is the angle of reflection from the grating:

$$sin\theta(\omega, k_x, k_y) = -\frac{2\pi}{k_{zx}}N + sin\left\{\alpha + atan\frac{k_x}{k_z}\right\} \qquad (3)$$

Hereinafter we assume the minus first diffraction order. In the chosen reference frame, the transverse wave vectors are related to the incidence angles $\alpha$ and $\gamma$ as $k_x = 0$, $k_y = \frac{\omega}{c}sin\gamma$. Taking into account the large beam size we neglect diffraction, i.e. the second derivatives of $\Psi$ with respect to $k_x, k_y$. Then, upon differentiation of (2) with allowance for (3) we find the derivatives of interest to us:

$$\Psi'_{k_x}\left(\omega, k_x = 0, k_y = \frac{\omega}{c}sin\gamma\right) = -X(\omega) = -L\frac{sin(\beta+\alpha)}{cos\beta} \qquad (4)$$

$$\Psi'_{k_y}\left(\omega, k_x = 0, k_y = \frac{\omega}{c}sin\gamma\right) = -Y(\omega) = -Ltan\gamma\frac{1+cos(\beta+\alpha)}{cos\beta} \qquad (5)$$

$$\frac{1}{2}\Psi''_{\omega\omega}\left(\omega = \omega_0, k_x = 0, k_y = \frac{\omega}{c}sin\gamma\right) = GVD = -\frac{L}{\omega_0 c}cos\gamma\frac{(sin\alpha-sin\beta_0)^2}{2cos^3\beta_0}, \qquad (6)$$

where the angle of reflection $\beta = \beta(\omega)$ is found from

$$sin\beta = -\frac{2\pi c}{\omega}\frac{N}{cos\gamma} + sin\alpha, \qquad (7)$$

and $\beta_0 = \beta(\omega_0)$. The expression for GVD (6) is derived in [21], and the expression (7) can be found in [22, 28]. The expression for $GVD$ (6) with allowance for (7) is the same as for $GVD$ for TC but with the substitution $L \to Lcos\gamma$; $N \to N/cos\gamma$. From (2, 3) it can be readily shown that this remark is true for all frequency derivatives, i.e. for all dispersion orders. This circumstance can be used, e.g., for a stretcher design.

We will consider only the case when the beam is not clipped on the second grating (the case of clipping was considered in detail in a number of works, e.g. [3, 5, 12, 29, 30], and will be briefly discussed in Section 4), so we will assume straight away that the beam size on the second grating coincides with its length $L_g$ and height $H_g$. Taking this into account, from Fig. 1 it can be found that

$$L_g = \frac{d+|X_b-X_r|cos\gamma}{cos\alpha} \qquad (8)$$

$$H_g = \frac{d+|Y_b-Y_r|cos\gamma}{cos\gamma} + (d - |X_b - X_r|)tan\gamma tan\alpha, \qquad (9)$$

where $X_b = X(\omega_b)$, $X_r = X(\omega_r)$, $Y_b = Y(\omega_b)$, $Y_r = Y(\omega_r)$, and $\omega_{b,r}$ are the high-frequency and low-frequency boundaries of the pulse spectrum. When deriving (8, 9), we took into account that $X_{b,r}$ and $Y_{b,r}$ are the beam coordinates in the plane perpendicular to the $z$ axis, but not in the plane normal to the beam, and also that the gratings are tilted in two planes (second term in (9)). From (8, 9) with allowance for (6, 4, 5) we obtain

$$L_g = \frac{d}{\cos\alpha} + L_{disp} \frac{2\cos^3\beta_0}{(\sin\alpha - \sin\beta_0)^2} |\tan\beta_b - \tan\beta_r| \tag{10}$$

$$H_g = d\left(\frac{1 + \tan\alpha \sin\gamma}{\cos\gamma}\right) + L_{disp} \frac{2\cos^3\beta_0 |\tan\gamma|}{(\sin\alpha - \sin\beta_0)^2} \left\{\frac{1}{\cos\gamma} \left|\frac{1+\cos(\beta_b+\alpha)}{\cos\beta_b} - \frac{1+\cos(\beta_r+\alpha)}{\cos\beta_r}\right| - \sin\alpha |(tg\beta_b - tg\beta_r)|\right\} \tag{11}$$

where $L_{disp} = |GVD|\omega_0 c$, and $\beta_b = \beta(\omega_b)$, $\beta_r = \beta(\omega_r)$. The absence of beam clipping along the $x$-coordinate leads to limitations on the beam size $d$, which follows from (10):

$$d < d_g = \left(L_g - L_{disp} \frac{2\cos^3\beta_0 \cos\gamma}{(\sin\alpha - \sin\beta_0)^2} |\tan\beta_b - \tan\beta_r|\right) \cos\alpha. \tag{12}$$

This expression is identical for TC and LC. In the $\omega_0 \gg \Omega$ approximation, (12) transforms to the expression obtained in [20] under this approximation. The second limitation on $d$ is the need to ensure decoupling of the beams, i.e. non-overlapping of the second grating with the incident beam. For TC, decoupling is attained in the direction of the $x$-axis (Fig. 1a). Obviously, for this the minimum beam displacement $|X_{min}|$ should be larger than the beam size $d$ plus the minimum required technological gap $g$:

$$|X_{min}| > d + g. \tag{13}$$

For $\alpha > \alpha_L$ (the case in Fig. 1a), $X_{min} = X_r$, and for $\alpha < \alpha_L$, vice versa, $X_{min} = X_b$. Taking this into account, from (4, 13) we obtain for TC the following expression

$$d < d_i = JL_{disp} - g \quad \text{(for TC)}, \tag{14}$$

where

$$J = \begin{cases} \frac{\sin(\beta_r+\alpha)}{\cos\beta_r} \frac{2\cos^3\beta_0}{(\sin\alpha-\sin\beta_0)^2} & \text{for } \alpha > \alpha_L \\ \frac{|\sin(\beta_b+\alpha)|}{\cos\beta_b} \frac{2\cos^3\beta_0}{(\sin\alpha-\sin\beta_0)^2} & \text{for } \alpha < \alpha_L \end{cases}. \tag{15}$$

The expression analogous to (14) was presented in [20] in different notation. For LC, decoupling occurs in the direction of the $y$-axis and requires that the minimum beam displacement $|Y_{min}|\cos\gamma$ should be larger than $(d+g)$. Since the gratings are tilted in two planes, then strictly speaking, $g$ is a function of the angles $\alpha$ and $\gamma$, but further for simplicity we will assume $g = const$. The most stringent condition for decoupling is for frequency $\omega_b$: $|Y_b|\cos\gamma > d + g$. With this taken into account, from (5) we obtain

$$d < d_i = IL_{disp} - g \quad \text{(for LC)}, \tag{16}$$

where

$$I = |\tan\gamma| \frac{1+\cos(\beta_b+\alpha)}{\cos\beta_b} \frac{2\cos^3\beta_0}{(\sin\alpha-\sin\beta_0)^2}. \tag{17}$$

In addition to meeting the conditions (12) and (14, 16), it is demanded that there be no diffraction orders other than the minus first one. This condition is always more stringent for radiation with frequency $\omega_b$. Let us introduce the function $\Pi(\alpha)$, which is equal to zero if at least one of these diffraction orders is

$$\Pi(\alpha) = \begin{cases} 0 & \text{if } \sin\alpha < 1 - \frac{2\pi c}{\omega_b}\frac{N}{\cos\gamma} \text{ or } \sin\alpha > \frac{4\pi c}{\omega_b}\frac{N}{\cos\gamma} - 1 \\ 1 & \text{if otherwise} \end{cases}. \tag{18}$$

The two conditions in the top line correspond to the first and minus second order of diffraction, respectively. Thus, the maximum beam size $D$, determined by simultaneous fulfillment of the three above conditions, has the form:

$$D = \min\{d_g; d_i\} \cdot \Pi(\alpha), \tag{19}$$

where $d_g$ and $\Pi(\alpha)$ are found from (12) and (18) for both compressors, and $d_i$ from (14) for TC and from (16) for LC. Note that the above expressions for LC are valid for any out-of-plane compressor, i.e. for any angle $\alpha$, as we have not used the condition $\alpha = \alpha_L$ when deriving these expressions.

It is convenient to conduct further discussion on the example of specific parameters of a compressed pulse, which will be addressed in the next two sections. Here, for reference we provide useful formulas for $L$ and $\alpha_L$ that follow from (6) and (7):

$$L = |\text{GVD}|\omega_0 c \frac{1}{\cos\gamma} \frac{2\cos^3\beta_0}{(\sin\alpha - \sin\beta_0)^2} \tag{20}$$

$$\sin\alpha_L = \frac{\pi c}{\omega_0} \frac{N}{\cos\gamma}. \tag{21}$$

## 3. Optimization of TC and LC for XCELS project

Let us consider the parameters for the XCELS project [4]: $L_g = 138$ cm, $\lambda_0 = 910$ nm, $\Delta\lambda = 150$ nm, $g = 5$ cm, and 2GVD=-4.42 ps². Here, 2GVD is the dispersion of two grating pairs, i.e. like above, GVD is the dispersion of one grating pair. As an example, the dependence of a number of parameters on $\alpha$ for $N = 1050/$mm is plotted in Fig. 2. The yellow line shows the Littrow angle for clarity. The green curve $d_g$ (12) corresponds to the restrictions on the beam size imposed by the condition of the absence of beam clipping. The blue curve $d_i$ corresponds to the restrictions on the beam size imposed by the need for decoupling in the diffraction plane for TC (14) (Fig. 1a) and in the orthogonal plane for the out-of-plane compressor (16) (Fig. 1b). The black meander shows the range of angles in which there are no other diffraction orders (18): the first order is possible to the left of the meander, and the minus second order to the right. Finally, the red dashed curve combines the three above restrictions for the $D(\alpha)$ relation (19). The maximum value of this curve corresponds to the maximum beam size (at $N$=1050/mm) and, therefore, the maximum output energy and pulse power after the compressor. The behavior of the curve $D(\alpha)$ greatly depends on $N$ for both TC and LC (Fig. 3). The curves in Fig. 3a (for TC) have two local maxima. At large $N$ the global maximum is at $\alpha > \alpha_L$, and at small $N$ at $\alpha < \alpha_L$.

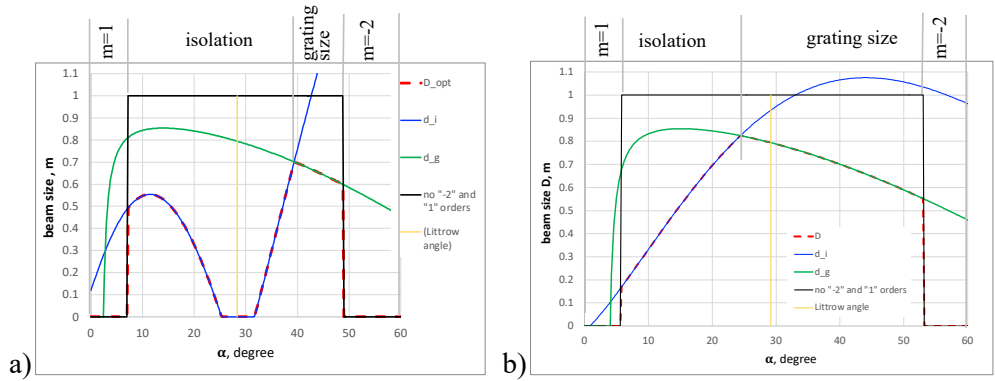

Fig. 2. Restrictions on maximum beam size at $L_g = 138$ cm, $N$=1050/mm for TC (a) and for out-of-plane compressor at $\gamma = 13°$ (b): green curve for $d_g$ (12) – no beam clipping on the grating; blue curve for $d_i$ (14), (16) – decoupling needed; black meander $\Pi(\alpha)$ (18) – range of angles without other diffraction orders (18): the first order is possible to the left of the meander and the minus second order to the right; red dashed curve combines all restrictions and shows $D(\alpha)$ (19); yellow line shows the Littrow angle for clarity.

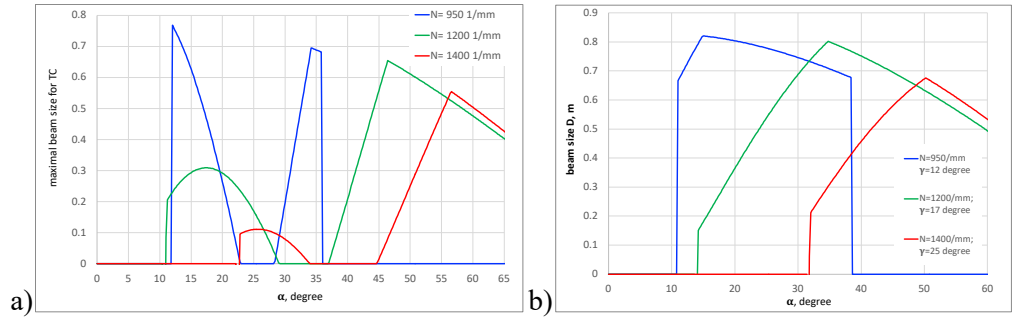

Fig. 3. Maximum beam size $D(\alpha)$ for TC (a) and for out-of-plane compressor (b) for $L_g = 138$ cm and $N$=950/mm (blue), $N$=1200/mm (green), $N$=1400/mm (red).

The parameters of an out-of-plane compressor can be optimized in a wide range of angles $\alpha$, including $\alpha < \alpha_L$. All the above expressions are valid for any $\alpha$. In what will follow we will restrict consideration to the case of LC ($\alpha = \alpha_L$, Fig. 1b) that is interesting from the practical point of view. Recently, the possibility of developing gratings having length $L_g = 160$ cm and parameters of a compressor with such gratings have been discussed in the literature [5, 16]. Here, we will find parameters of the optimal compressor for XCELS for two options: $L_g = 138$ cm and $L_g = 160$ cm.

The maximum size of the beam $D$, both in LC and TC, depends on two parameters: $N$ and $\alpha$ for TC and $N$ and $\gamma$ for LC. For each $N$ there exists an optimal value of the angle $\alpha_{opt}$ or $\gamma_{opt}$ at which $D$ has a maximum. The relations $D_{opt}(N) = D(N, \alpha_{opt})$ for TC and $D_{opt}(N) = D(N, \gamma_{opt})$ for LC are shown in Fig. 4a,b by triangles for $L_g = 138$ cm and by squares for $L_g = 160$ cm. For $L_g = 138$ cm, the maximum value of the beam size $D_m$ is a little larger for LC: 79 cm versus 78 cm. For TC, $D_{opt}(N)$ has a well pronounced maximum at $N = 950$/mm, whereas for LC, conversely, a plateau in the $N = (1000 \ldots 1250)$/mm range. This is an advantage of LC, since it gives freedom to choose $N$. The choice of a specific value of $N$ may be made, for example, for reasons of a higher efficiency, a higher laser induced damage threshold of the grating, etc. Note that $D_m$=70 cm is much larger than the beam diameter in the initial XCELS design (see Table 1). An analogous plateau in the $N = (950 \ldots 1150)$/mm range is observed in the $D_{opt}(N)$ function for LC at $L_g = 160$ cm. In this case, LC is obviously more preferable, since it enables a larger value of $D_m$: 97 cm versus 86 cm for TC.

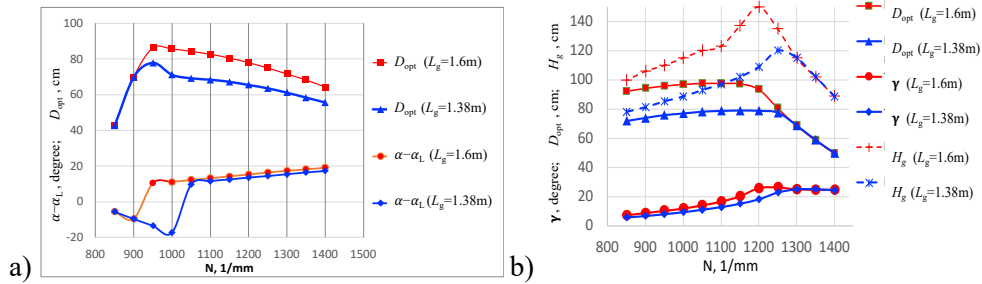

Fig. 4. Curves for compressor parameters for XCELS for TC (a) and LC (b) with grating length $L_g = 138$ cm (blue) and $L_g = 160$ cm (red). Squares and triangles – beam size $D_{opt}$ at optimal angles $\alpha$ and $\gamma$, circles and diamonds – difference between incidence angle in the diffraction plane and Littrow angle ($\alpha - \alpha_L$) (a) and incidence angle in the plane orthogonal to the diffraction plane $\gamma$ (b); plus signs and asterisks (b) – grating height $H_g$.

Table 1. Compressor parameters

|  | XCELS $\lambda = (910 \pm 75)$ nm | | | | | SEL=100 PW $\lambda = (925 \pm 100)$ nm | |
|---|---|---|---|---|---|---|---|
|  | $L_g = 138$ cm | | | $L_g = 160$ cm | | $L_g = 160$ cm | |
|  | TC ([4]) | TC (new) | LC | TC | LC | TC | LC |
| $N$, 1/mm | 1200 | 950 | 1100 | 950 | 1000 | 1000 | 1100 |
| $\alpha$, degree | 46.2 | 12.2 | 30.6 | 36.0 | 27.3 | 38.8 | 31.6 |
| $\gamma$, degree | 0 | 0 | 12.9 | 0 | 9.4 | 0 | 15.4 |
| $D_m$, cm | 66 | 78 | 79 | 86 | 97 | 75 | 87 |
| $H_g$, cm | 66 | 78 | 97 | 86 | 111 | 75 | 115 |
| $W^a$, J | 1006 | 1400 | 1440 | 1720 | 2170 | 1284 | 1730 |
| $\tau$, fs | 20 | 20 | 20 | 20 | 20 | 15 | 15 |
| $P$, PW | 50 | 70 | 72 | 86 | 109 | 86 | 115 |

a) given that $R\eta w_{th} = 0.231$ J/cm$^2$ in the plane normal to the beam

The circles and diamonds in Fig. 4a correspond to the dependence of $(\alpha - \alpha_L)$ on $N$. It is clearly seen that for large $N$, $\alpha > \alpha_L$, which corresponds to a standard compressor design for high-power lasers. At the same time, for small $N$, maximum beam size $D_m$ is attained at $\alpha < \alpha_L$. This is also well seen in Fig. 3a (left maximum in the blue curve above the right maximum). We are not aware of the usage of TC with $\alpha < \alpha_L$ in high-power lasers. The circles and diamonds in Fig. 4b show $\gamma(N)$ at which $\alpha = \alpha_L$. In the region of the $D_{opt}(N)$ plateau, i.e. at $N = (1000 \ldots 1200)$/mm, $\gamma = 10° \ldots 20°$, which falls within the range where the efficiency of the gratings almost does not decrease [16, 24].

It is worth noting the LC drawback: the grating height $H_g$ is larger than the beam size. The dashed curves in Fig. 4b show $H_g(N)$ plotted by the expression (11). At the same time, the increase in $H_g$ required for LC is not so great – compare the curves for $H_g(N)$ and $D_{opt}(N)$, and may well be implemented in practice. Still another LC drawback is that in a general case the choice of input beam polarization is nontrivial. This issue was studied in detail in [24]. From the analysis made in [24] it follows that vertical incident polarization, when the field is normal to the direction of the grooves, is optimal (Fig. 1b). The experiment [24] carried out at $\gamma = 15°$ showed that in this case the reflection coefficient of one grating $R$ and the efficiency of the entire compressor differ negligibly from the corresponding parameters at $\gamma = 0$. These results were obtained for a wavelength of 800 nm and $N = 1480$/mm; they need clarification for other wavelengths and groove densities.

The main parameters of TC and LC for the XCELS project are presented in Table 1. For comparison of different designs, it also contains values of maximum beam energy $W$ calculated by the expression (1), given that $R\eta w_{th} = 0.231$ J/cm$^2$, which corresponds to $R = 0.92$ and to the value of safe fluence $\eta w_{th} = 0.251$ J/cm$^2$ in the plane normal to the beam, i.e. 0.174 J/cm$^2$ on the grating surface at $\alpha = 46°$ [4]. Note that this is a rather conservative estimate, since gratings with $w_{th} = 0.4$ J/cm$^2$ and $w_{th} = 0.57$ J/cm$^2$ in the plane normal to the beam are reported in [31] and [16], and $\eta = 1.31$ [4, 10] or $\eta = 1.41$ [5] are considered in the literature for $\eta$. The FTL pulse in XCELS has a duration of 17 fs, whereas the values of maximum power in Table 1 are given for a 20 fs pulse that is more real in practice. It is clear from the table that the new TC and LC designs with a grating length of 138 cm allow increasing the output power by a factor of 1.4 and 1.44, i.e. up to 70 PW and 72 PW, respectively. With the use of 160x111-cm gratings in LC, over 110 PW may be achieved.

## 4. Compressor for SEL-100 PW

Let us consider the parameters for the SEL-100 PW project [1, 3, 10]: $L_g = 160$ cm, 2GVD=-4.6 ps$^2$, $\lambda_0 = 925$ nm, $\Delta\lambda = 200$ nm, and $g = 5$ cm. For these values, the optimal parameters for TC and LC are listed in Table 1. Since the pulse spectrum width in the SEL-100 PW is 1.33 times larger than in XCELS, for a correct comparison we assume the 15-fs pulse duration to be 1.33 times shorter than in XCELS. It is seen from the table that, for a grating size of 160x75 cm, the optimal design of the TC provides an output power of 86 PW. In this case, the angle of incidence $\alpha$ differs from the Littrow angle only by 11.5 degrees. LC allows achieving a significantly higher power of 115 PW with 160x115-cm gratings. The angle $\gamma$ in this case, despite being larger than in the other designs presented in Table 1, still falls within the range in which the grating efficiency almost does not reduce [16, 24].

It is important to note that the analysis made in this work completely excludes beam clipping by gratings. The design of the two-grating compressor for the SEL-100 PW presented in [5] implies strong clipping. This leads to three effects that reduce the focal intensity: pulse stretching due to narrowing of the spectrum, loss of radiation energy, and deterioration of focusability. In the example numerically calculated in [5], the losses were approximately 11%, 7.8% and 15%, i.e. more than 35% in total. It is worthy of note that these losses cannot be compensated by increasing pulse energy at the compressor input, as clipping does not reduce fluence on the last grating. Therefore, according to (1) the compressor [5] enables 35% lower focal intensity than a compressor without clipping for the same values of $w_{th}$, $R$, $\eta$, and $d$.

Comparison of the compressor parameters for XCELS and SEL-100 PW with 160-cm long gratings shows that for TC the maximum achievable power is the same – 86 PW; whereas for LC the SEL-100 PW power is 5.5% higher – 115 PW versus 109 PW. However, from a practical point of view, the XCELS option is preferable, since for a narrower pulse spectrum, the requirements for both the compressor gratings and the rest of the optics are lower. At the same time, XCELS requires 1.33 times higher pulse energy, hence, DKDP crystals with $\sqrt{1.33} = 1.15$ times larger size are required.

All compressor variants discussed above are symmetric: $L_2 = L_1$; $N_2 = N_1$; $\alpha_2 = \alpha_1$; $\gamma_2 = \gamma_1$, where the indices "1" and "2" correspond to the first and second grating pairs. At the same time, they can be easily modified into asymmetric compressors that ensure reduction of fluence fluctuations due to the time delay of high-frequency spatial harmonics [13, 15] or spatial dispersion of the output beam [5, 12, 14]. In asymmetric compressors, grating pairs differ from each other: $L_2 \neq L_1$ [5, 12, 14]; $N_2 \neq N_1$, $\alpha_2 \neq \alpha_1$ [13]; $\gamma_2 \neq \gamma_1$ [15]. Note that $\gamma_2$ and $\gamma_1$ can have not only different absolute values, but also signs. For example, LC with $\gamma_2 = -\gamma_1$, in which gratings of the first and second pairs are parallel in the y-plane and antiparallel (mirror) in the x-plane, are 2 times shorter and 2 times wider than for the case $\gamma_2 = \gamma_1$. For $\gamma_1 \approx 10°$, fluence fluctuations are radically suppressed. The drawback of such a compressor is an additional increase in the grating height $H_g$.

## 5. Conclusion

Since in high-power femtosecond lasers the output pulse energy is limited by the laser induced damage threshold of the last diffraction grating of the compressor, the optimal compressor design is the one ensuring maximum size of the output beam. For given parameters of a chirped pulse (central frequency, bandwidth, GVD) and a given diffraction grating length $L_g$, an analytical expression has been obtained for the maximum beam size $D$, at which both beam clipping on the gratings and the appearance of spurious diffraction orders are completely excluded. Using this expression, it is easy to find optimal compressor parameters that allow obtaining maximum $D$: the distance between the gratings along the normal $L$, the groove density $N$, the angle of incidence on the first grating in the diffraction plane $\alpha$, and the angle of incidence on the first grating outside the diffraction plane $\gamma$.

The analysis was performed in a general form for an out-of-plane compressor, i.e. for arbitrary values of the angles $\alpha$ and $\gamma$. Two particular "plane" cases attractive for practical reasons were considered: a standard Treacy compressor ($\gamma = 0$, Fig. 1a) and a Littrow compressor with an incidence angle equal to the Littrow angle ($\alpha = \alpha_L$, Fig. 1a). The Littrow compressor almost always ensures a larger value of $D$ than the Treacy compressor. For the TC compressor, $D(N)$ has a well pronounced maximum determining the choice of $N$ (Fig. 4a). For LC, $D(N)$ has a form of a plateau (Fig. 4b), which allows choosing $N$ within this plateau for technological reasons: the larger the reflection coefficient, the higher the laser damage threshold.

Optimal TC and LC designs that enable a substantial output power increase (by tens of percent) were calculated for the pulse parameters of the XCELS and SEL-100 PW projects. In particular, for 160-cm long diffraction gratings, the optimal TC design allows obtaining 86 PW for both projects, and the optimal LC design 109 PW and 115 PW for XCELS and SEL-100 PW, respectively.

## Acknowledgments

The work was supported by the Ministry of Science and Higher Education of the Russian Federation (075-15-2020-906, Center of Excellence "Center of Photonics"). Author thanks Anton Vyatkin and Ivan Yakovlev for the fruitful discussions.

## Disclosures

The authors declare no conflicts of interest.

## Data Availability

Data underlying the results presented in this paper are not publicly available at this time but may be obtained from the authors upon reasonable request.